**GNC Analysis and Robotic Systems Configuration of Collision-free Earth Observation Satellites (CfEOS) Constellations**

*Manuel Ntumba [(1)], Saurabh Gore [(2)], Pulkit Jain [(3)], Jean-Baptiste Awanyo [(4)]*

[(1)(3)] *Texas Instruments Innovation Lab, Department of Mechatronics, Chandigarh University, India.*

[(2)] *Department of Aerospace Engineering, Moscow Aviation Institute, Moscow, Russia.*

[(4)] *Univerité Sultan Moulay Slimane, Béni Mellal, Morocco.*

[(1)] *Space Generation Advisory Council, Vienna, Austria.*

manuel.ntumba@spacegeneration.org

sdgore@mai.education

**Abstract**

The high number of objects in the LEO is a risk that collisions between sub-orbital or escape velocity objects with an orbiting object of satellites occur when two satellites collide while orbiting the earth. One of the approaches to avoid collisions is a robotic configuration of satellite constellations. Satellite constellations should not be confused with satellite clusters, which are groups of satellites moving in close proximity to each other in nearly identical orbits; nor with satellite series or satellite programs, which are generations of satellites launched successively; nor with satellite fleets, which are groups of satellites from the same manufacturer or operator that operate an independent system. CfEOS constellations designed for geospatial applications and Earth observation. Unlike a single satellite, a constellation can provide permanent global or near-global coverage anywhere on Earth. CfEOS constellations are configured in sets of complementary orbital planes and connect to ground stations located around the globe. This paper describes the GNC analysis, the orbit propagation and robotic systems configuration for Collision-free Earth observation satellites (CfEOS) constellations.

1. **Introduction**

The satellites are maneuvered from the ground station to avoid potential collisions. For instance, in May 2013, NEE-01 Pegaso CubeSat encountered physical damage and loss of communication after a close encounter of particles from a debris cloud around an upper stage of Tsyklon -3 (SSC-15890) left by launch of Kosmos 1666 into low earth orbit (LEO). [2] [4] Satellite constellations in LEO have lower path losses, lower costs, and low latency communication. Walker constellations are based on a simple design approach to assigning satellites in a constellation based on Ascending Node Longitude or RAAN: Right Ascension of Ascending Node. [1] [3] The sequence of commands, to reposition moving parts, such as antennas and solar panels or sensors, to collect and report spacecraft telemetry data to the correct orientation of the spaceship in space despite the external effects of the gravity gradient. [5] [6] The GNC system and robotic setup are responsible for maintaining magnetic fields, solar radiation and aerodynamic drag, and reporting mission data to optimize parameters. [7]

## 2. Mission Concept

In order to provide global coverage for Earth observation purposes, CfEOS constellations using a number of different orbital planes might be possible. Usually, constellations are designed taking into account that satellites will have similar orbits, eccentricity, and inclination, to implement the same disturbance model and corrective maneuvers for each satellite from an individual orbit. The advantage of using circular orbit is that they provide a constant altitude which ensures a constant signal strength to communicate with the ground stations. Another important aspect when designing the constellation is the phasing of each satellite in an orbital plane to maintain sufficient separation and avoid collisions or interference at intersections of the orbital plane. Based on this notation, our constellation model can be represented by 55: 36/6/60, which has a semi-major axis of 7170 km, the eccentricity is 0 (circular orbit), an inclination of 55 degrees. Six orbital planes inclined at 55 degrees contain 6 equidistant satellites in each orbit forming a network of 36 satellites in low earth orbit. Three satellites (Comms-1, Comms-2 and Comms-3) are in equatorial orbit for communication and telemetry, confirming access to all satellites once per day for data collection.

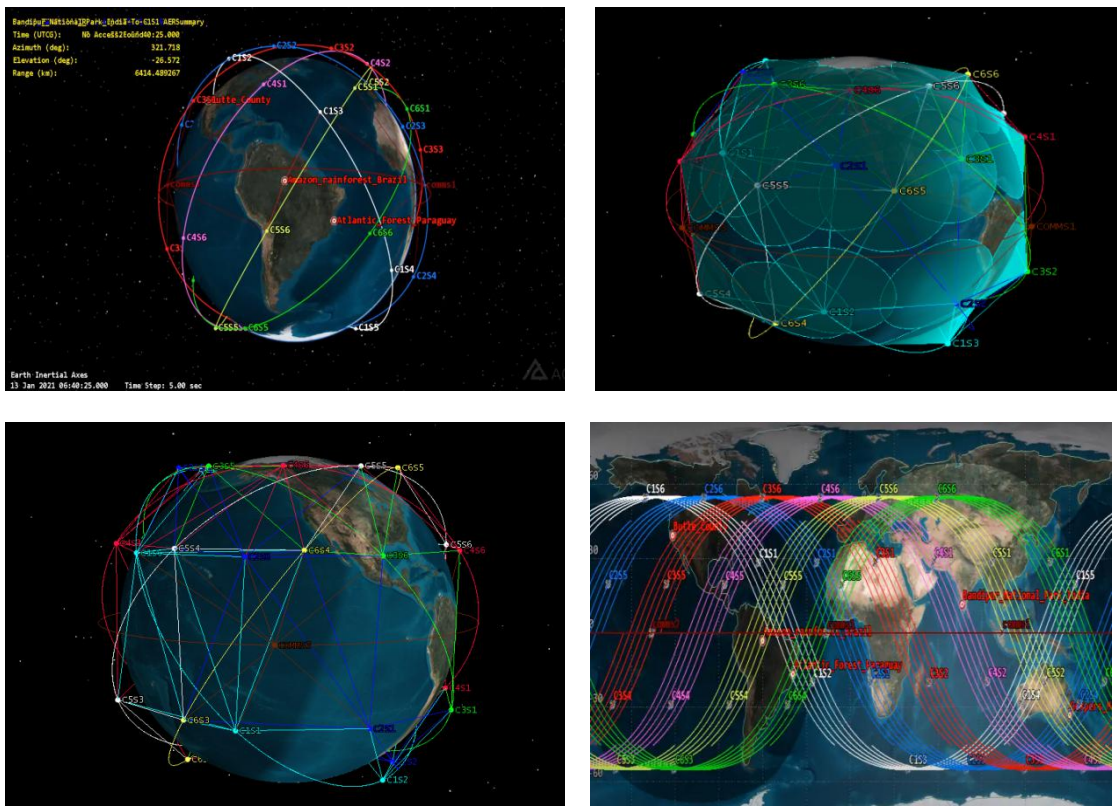

**Figure 1. Digital Mission Analysis, Data Transmission and Crosslink for CfEOS constellations**

## 3. Robotic Systems Configuration and GNC Analysis: Guidance, Navigation and Control

As the propagation of a multi-body dynamic takes into account the control of the robotic arm, the algorithms implemented are integrated into the CfEOS constellations. The reception and processing of the incoming signal are common to all receivers; on the other hand, the position calculation is carried out only by the CfEOS Constellations receivers. The CfEOS Constellations provides a basic receiver model that can calculate the number of visible satellites that can be tracked over time outage. The navigation function of the CfEOS Constellations is used to determine the accuracy that could not be achieved if a satellite were navigated only. The use of such a technique on CfEOS Constellations to be used for coverage on the Earth's surface. According to the GNC analysis, the configurations of the receivers and the signals, make it possible to navigate the CfEOS Constellations. The navigation analysis evaluates the link between the selected receptors and the CfEOS Constellations.

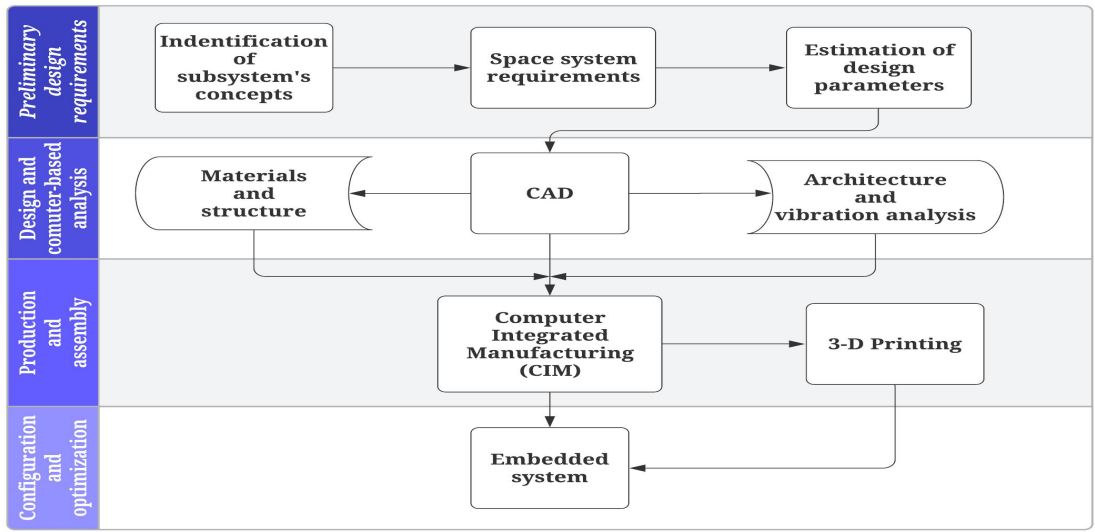

**Figure 2. Manufacturing Processes for CfEOS constellations**

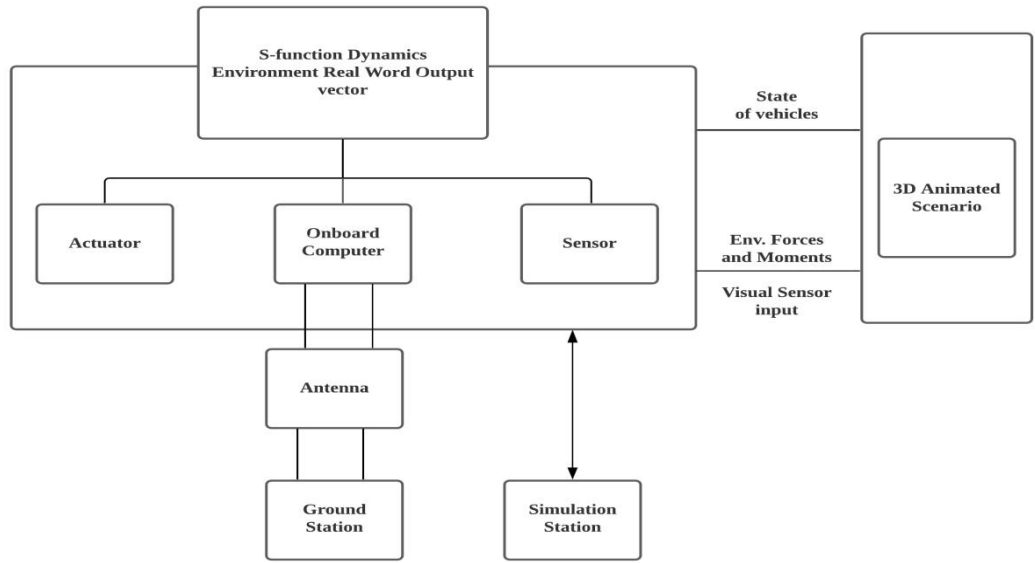

**Figure 3. Numerical Processes for CfEOS constellations**

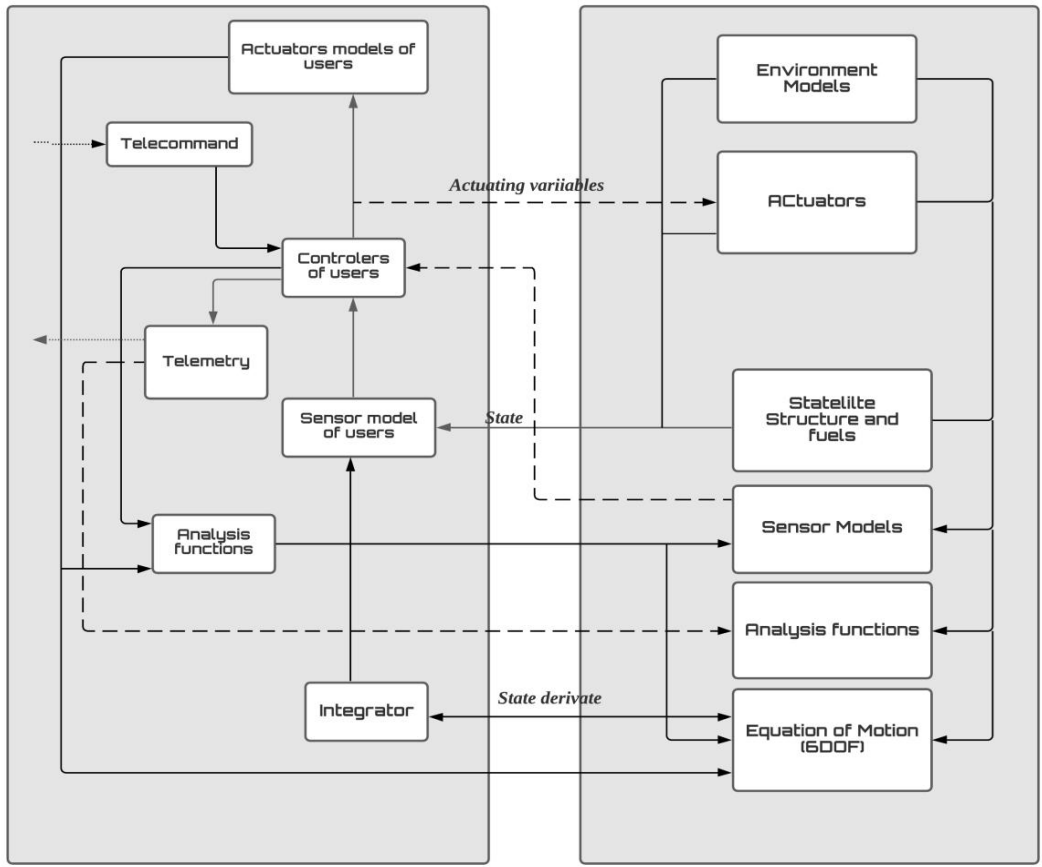

Figure 4. Robotic Systems Configuration for CfEOS constellations

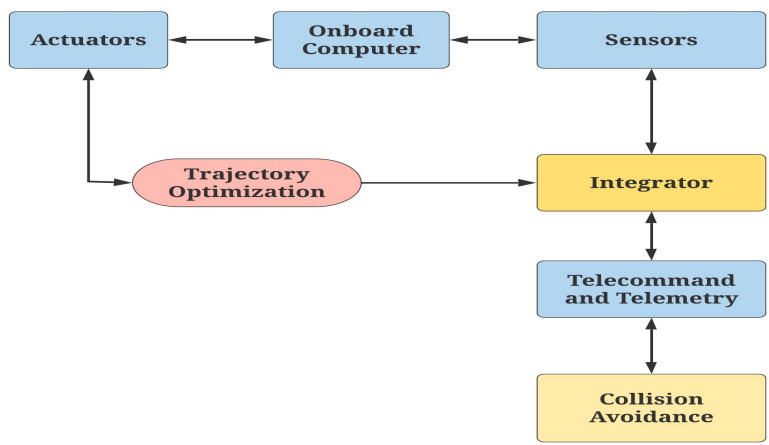

Figure 5. GNC Configuration for CfEOS constellations

CfEOS Constellations are integrated with communication terminals which include radio antennas, transmitters, and receivers; which allows communication with ground stations or with other spacecraft. The power configuration for CfEOS Constellations comes from photovoltaic solar cells or a radioisotope thermoelectric generator, batteries, and distributed circuits. CfEOS Constellations are protected against debris, solar heating and from temperature fluctuations by insulation. [8][9][10] Such shielding effects could obviously also become relevant in communication with ground stations or relay satellites. In addition, the performance of CfEOS constellations depends on visual sensors, data processing, and data transfer. It is possible to analyze the operating modes of the sensors. Such sensors are essential for orbital rendezvous or maneuvers. [11][12]

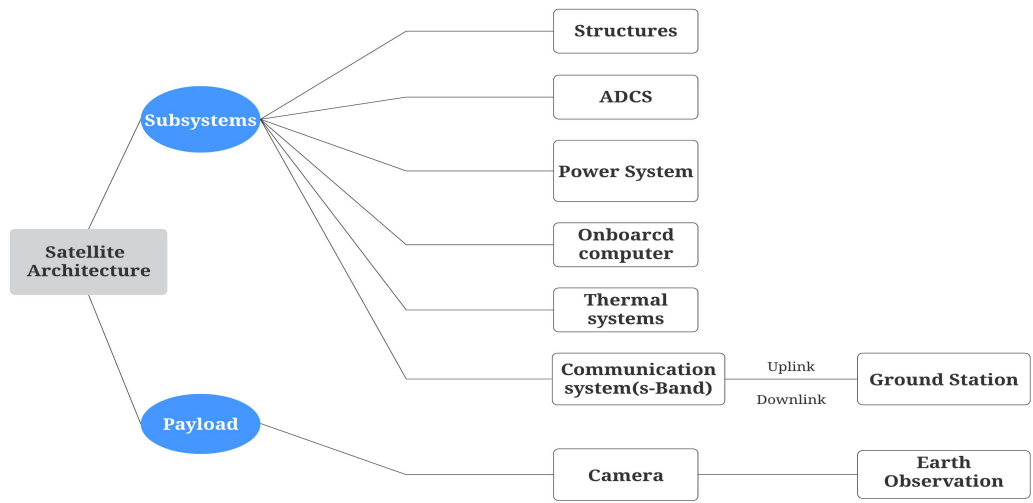

Figure 6. Architecture and Systems Configuration for CfEOS constellations

### 4. Data Link and Simulation Results

The link budget via the CfEOS Constellations, which provides autonomous geospatial positioning with global coverage. The robotic systems configuration supports the GNC and the other control systems of the CfEOS Constellations. CfEOS constellation subsystem models are refined by exchanging simple parameter definitions with more complex algorithms. The robotic configuration of the CfEOS constellations conditions the ascent of the launcher, reentry, and rendezvous by carrying out the entire GNC design for the aiming of the sensors inside the CfEOS constellations. [13][14] The performance of CfEOS constellations depends on the dynamics of the spacecraft and its interaction with the environment, the optimization of trajectories, orbit propagation, impulsive maneuvers, or optimal control. Mission analysis of CfEOS constellations can be performed by analyzing delta-V budget, ground station visibility including the selection of communication networks, visibility of other satellites. Relay communication or navigation, the field of view of navigation and payload sensors, the maximum volume of transferable data sets a gradual improvement in mission analysis by examining the pointing of sensors and antennas without ignoring the attitude of the CfEOS constellations. [15][16]

| Comms-1 to satellite access(sec) | Comms-2 to satellite access(sec) | Comms-3 to satellite access(sec) |
|---|---|---|
| C1S3 - 466.630 | C1S1 - 466.507 | C1S1 - 466.507 |
| C1S4 - 2768.377 | C1S5 - 466.621 | C1S2 - 2770.415 |
| C1S5 - 466.507 | C1S6 - 2770.415 | C1S3 - 466.485 |
| C2S3 - 2310.157 | C2S5 - 2312.225 | C2S1 - 2310.682 |
| C2S4 - 1606.677 | C2S6 - 1603.680 | C2S2 - 1602.560 |
| C3S2 - 1601.642 | C3S4 - 1606.755 | C3S1 - 2310.915 |
| C3S3 - 2310.042 | C3S5 - 2312.111 | C3S6 - 1602.560 |
| C4S1 - 466.0 | C4S3 - 466.765 | C4S1 - 466.377 |
| C4S2 - 2768.108 | C4S4 - 2761.408 | C4S5 - 466.756 |
| C4S3 - 466.356 | C4S5 - 466.377 | C4S6 - 2770.146 |
| C5S1 - 2310.192 | C5S3 - 2312.26 | C5S5 - 2312.260 |
| C5S2 - 1601.550 | C5S4 - 1606.663 | C5S6 - 1603.665 |
| C6S1 - 2317.302 | C6S2 - 1603.823 | C6S4 - 1606.821 |
| C6S6 - 1606.821 | C6S3 - 2312.239 | C6S5 - 2312.239 |

**Table 1. Communication Access Summary for CfEOS constellations**

The CfEOS constellations link budget analysis defines the signal-to-noise ratio as well as the link budget between a satellite and a ground station or two satellites. It takes into account the characteristics of the antenna of the receiver and transmitter as well as atmospheric effects. The data communication system replicates the data cycle by sensors on storage devices to ground contact and signal-based transmission to the ground. CfEOS constellations operate through the data rate of the sensors as a function of the mission time and to analyze the timing of the data transfer, signal loss during transfer. [17] CfEOS constellations included a high degree of dependencies between subsystems and analytical procedures. The GNC analysis provides real-time data such as differential forces, moments due to solar pressure, air drag which are sent back to the CfEOS constellations to avoid collisions. CfEOS Constellations are systems specially designed for a specific hostile environment thanks to an optimal amount of thrust produced. Performance of CfEOS Constellations depends on their specifications for a particular environment and their complexity. CfEOS constellations use telemetry to share acquired data between them and with the ground station. [18][19]

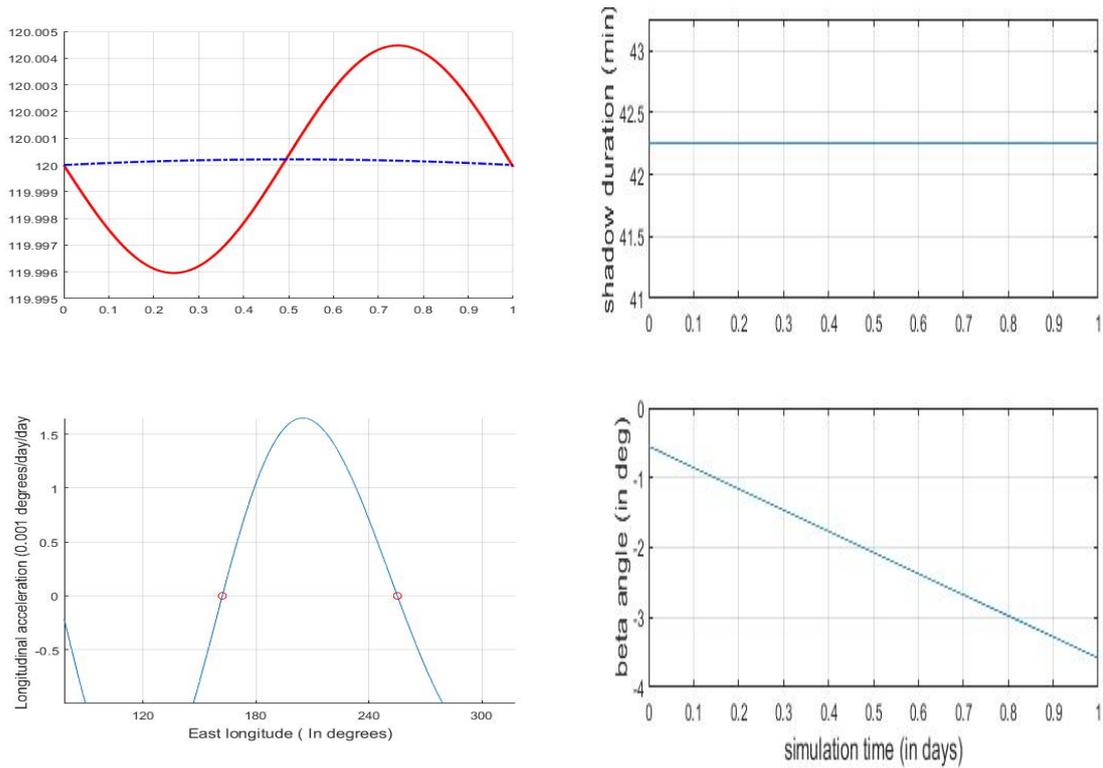

**Figure 7. Numerical Analysis for CfEOS constellations**

CfEOS constellations must also effectively perform risk assessment and real-time trajectory adjustments to avoid risks. To achieve this, they require precise knowledge of the spacecraft's location relative to the surface, what can present terrain hazards, and the current orientation of the CfEOS constellations. Without the capability of locate, risk assessment and avoidance operations, the CfEOS constellations become dangerous and can easily enter dangerous situations such as surface collisions, unwanted fuel consumption levels and dangerous maneuvers. Built-in sensing incorporates an image transformation algorithm to interpret immediate terrestrial imagery data, perform real-time detection, and avoid terrain hazards that can hinder a safe landing and increase the accuracy of landing on the ground. The integrated detection systems of the CfEOS Constellations complete these tasks by relying on pre-recorded information and the cameras are also used for risk assessment, detecting any possible danger, whether it is an increase in fuel consumption.

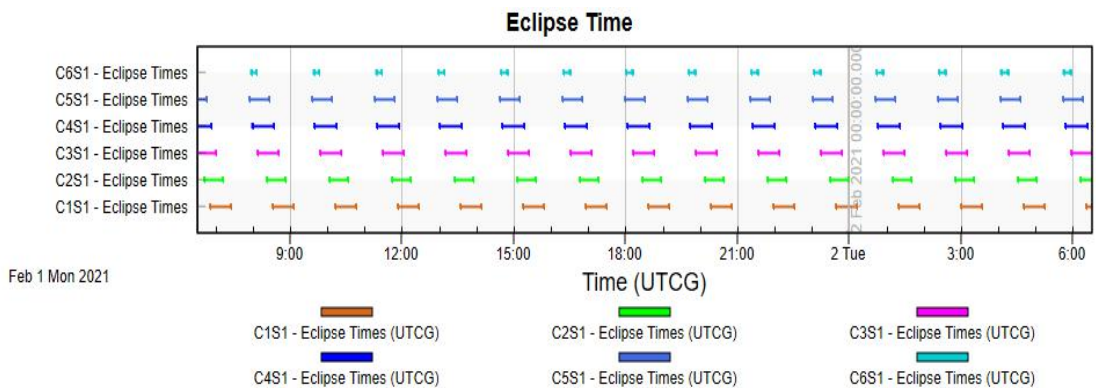

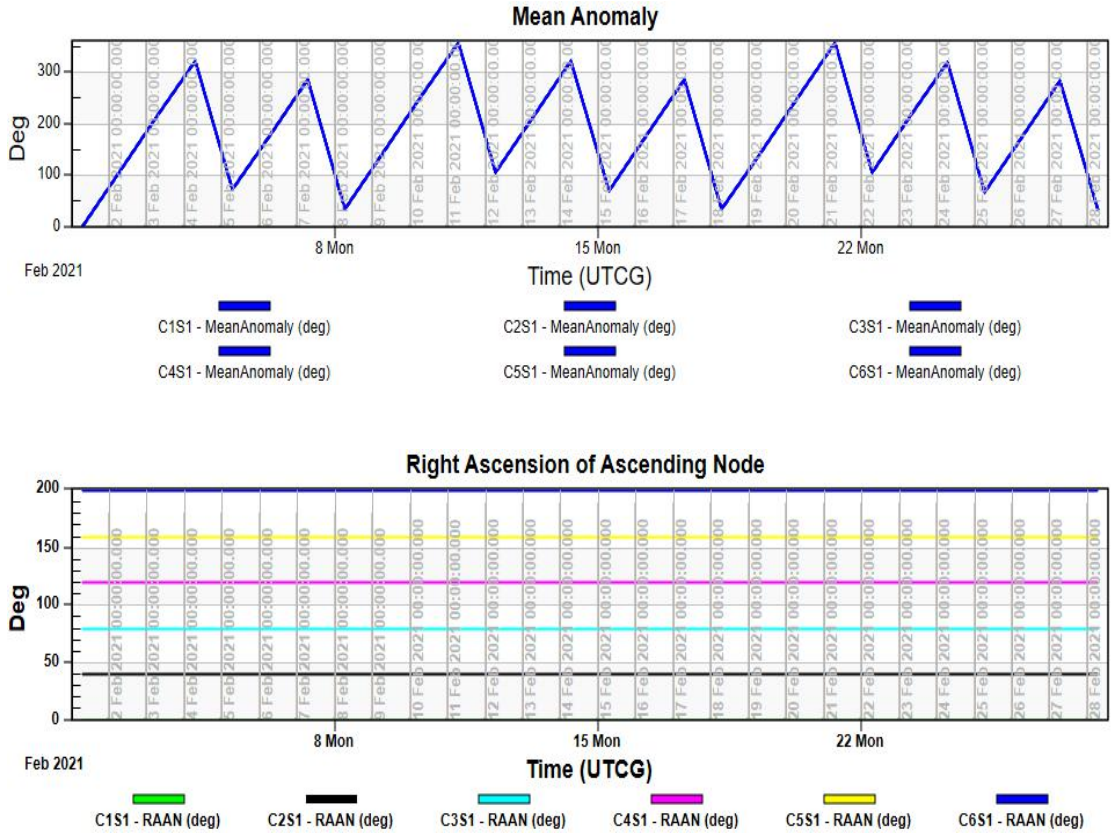

**Figure 8. Orbit Determination and Propagation for CfEOS constellations**

## 5. CONCLUSION

In this paper, we showed the complementarity between the robotic configuration, the link budget, the navigation, the power, the thermal management, the close approach maneuvers, and the coupled analysis of CfEOS constellations. Trajectory determination, ADCS analysis, and GNC analysis, and robotic configuration on CfEOS constellations provide the ability to compare specification values with simulation results. The robotic configuration of CfEOS constellations define reliable and suitable subsystem models over the lifetime of the satellites. It is therefore necessary to carry out automatic processing of the control cycles, of the design process, and of the trajectory planning. In particular, the multi-vehicle benefits from the coupled GNC analysis on CfEOS Constellations which reduces the risk of cost overruns and reduces the manufacturing time. A collision-free constellation is obtained through a crosslink architecture and inter-satellite communication.